\documentstyle[twoside,fleqn,espcrc2]{article}

\newcommand{\be}{\begin{eqnarray}}
\newcommand{\ee}{\end{eqnarray}}
\newcommand{\ben}{\begin{eqnarray*}}
\newcommand{\een}{\end{eqnarray*}}
\newcommand{\la}{\left\langle}
\newcommand{\ra}{\right\rangle}

\def\lab#1	{\hbox{\small #1} }
\def\mb#1       {\mbox{\boldmath $#1$}}
\def\smprod	{\mathop{\textstyle \prod}\limits}

\title{%
	\vspace{-10ex}
	{\small
	\begin{tabbing}
	\` {\sl hep-lat/9709084} \\
    	\\
	\` LSUHE No. 266-1997 \\
	\` September, 1997 
	\end{tabbing}
	}
 	Ehrenfest theorems for field strength and electric current in
	Abelian projected SU(2) lattice gauge theory}
\author{%
	Giuseppe DiCecio\thanks{%
	Present address: 
   	Dept. of Physics, Univ. of Pisa, Italy.},  
        Alistair Hart and Richard W. Haymaker\thanks{%
	Presented by R. Haymaker.
  	Work partially supported by the U. S. Department of Energy.}    
\\
\vspace{1ex}
        Department of Physics and Astronomy, 
       	Louisiana State University, Baton Rouge, Louisiana
       	70803,  USA}        
       	
\begin{document}

\begin{abstract}
	We derive an Ehrenfest theorem for SU(2) lattice gauge theory
	which, after Abelian projection, relates the Abelian field
	strength and a dynamical electric current and defines these
	operators for finite lattice spacing. Preliminary results from
	the ongoing numerical test of the relation are presented,
	including the contributions from gauge fixing and the
	Faddeev--Popov determinant (the ghost fields) in the maximally
	Abelian gauge.
\end{abstract}

\maketitle

	After applying Abelian projection to non--Abelian gauge
	theories, the result is mathematically identical to a set of
	charged fields (vector--like in the maximally Abelian (MA)
	gauge) coupled to an electromagnetic field, governed by a
	complicated U(1) gauge invariant action.  The charged fields
	have traditionally been ignored, but numerical work suggests
	that we should re-examine their r\^ole.

	We are presently completing a numerical check of an exact
	relation for SU(2) lattice averages after Abelian projection
	to general gauges,
\be
\la \Delta_{\mu} F_{\mu \nu} - J_{\nu} 
\ra_{\lab{static Abelian source} } = 0.
\label{eqn_ehren}
\ee
	$J_{\nu}$, the electric current in the remaining U(1),
	contains terms from the external source, the ``off-diagonal''
	gauge potentials, the gauge fixing condition and
	the Faddeev--Popov determinant:
\ben
J_{\nu} &=& J_{\nu}^{\lab{(source)} } +  J_{\nu}^{\lab{(dynamical)} } \\
&+& 
J_{\nu}^{\lab{(gauge fixing)} } +  J_{\nu}^{\lab{(Faddeev--Popov)} }.
\een
	Eqn. ~(\ref{eqn_ehren}) resembles the Euler-Lagrange--Maxwell
	equation, of course, satisfied at the extremum of the action.
	With a suitable choice of lattice operators, however, lattice
	averages also satisfy this relation.  The corresponding
	relations for U(1) and for SU(3) without gauge fixing are
	given in Ref.
\cite{zach}.
	The term ``Ehrenfest theorem'' is taken from the context of
	quantum mechanics, where a classical equation is exactly
	satisfied for expectation values, e.g.  $ \frac{d^2}{dt^2}\la
	\mb{r} \ra = - \la \mb{\nabla} V(\mb{r} ) \ra$.

\subsection*{\bf Motivation I:  
	Eqn.~(\ref{eqn_ehren}) defines electric current density on
	the lattice.}

\begin{itemize}
\item   
	Unlike pure U(1), an electric current density occurs in
	the Abelian projected theory, and is capable of screening
	sources and affecting the string tension; Bali et. al.
\cite{bbms} 
	found that the string tension after Abelian projection to the
	MA gauge is only $92\%$ of the full SU(2) result.
\item 
	The contributions from gauge fixing and the Faddeev--Popov
	determinant contribution
\cite{thanks} 
	to the current can be measured by making use of this
	relation. The latter is the contribution from the
	ghost fields.
\item 
	One uses the Ginzburg--Landau theory interpreted as a dual
	effective theory to model the simulations. One needs to modify
	this model, however, to include the effects of a dynamical
	charge density.
\end{itemize}

\subsection*{\bf 
	Motivation II: Eqn.~(\ref{eqn_ehren}) defines Abelian field
	strength on the lattice.}

\begin{itemize}
\item 
	For regions where $J_{\nu}= 0$, this defines exactly the
	conservation of flux.  It then gives precise meaning to the
	intuitive picture of the vacuum squeezing the field lines.
\item 
	Crucial to this mechanism for confinement is the connection
	between the spontaneous breaking of a gauge symmetry as
	indicated by a non-zero vacuum expectation value of the
	monopole creation operator and the formation of vortices.
	Both the monopole operator
\cite{ddpp,cpv1} 
	and the vortex operators
\cite{sbh} 
	rely on a definition of electric field strength.  Therefore a
	tightening of these definitions could enhance our
	understanding of this crucial connection.
\item  
	One can compare this definition of Abelian field strength with
	the lattice implementation
\cite{bdh} 
	of the 't Hooft expression
\cite{thooft2} 
	which would lead to an exact Ehrenfest theorem only to leading
	order in the lattice spacing, $a$.
\end{itemize}

\subsection*{\bf U(1) Example}

	Consider the partition function for the U(1) plaquette
	action, 
$S_a = \sum_{n,\mu,\nu} (1-\cos \theta_{\mu \nu}(n))$, 
	including a Wilson loop:
\ben
{\cal Z}_W = \int [d \theta_{\mu}] e^{i \theta_W} 
\exp{\left( -\beta S_a \right)}.
\een
	It is invariant under shifting any link angle,
	$\theta_{\mu}(n) \rightarrow \theta_{\mu}(n) +
	\epsilon$. Using this invariance Zach et. al.
\cite{zach}
	derived the relation:
\ben
\frac{ 
\int [d \theta_{\mu}] \sin{ \theta_W} 
\left( \frac{1}{a} \Delta_{\mu} \frac{\sin \theta_{\mu \nu}}{e a^2} \right)
e^{-\beta S_a}
}
{ 
\int [d \theta_{\mu}] \cos{ \theta_W} 
e^{-\beta S_a}
}
= e \frac{\delta_W}{a^3}
\een
	where $\delta_W= 0$ unless the shifted link lies on the Wilson
	loop external source, when $\delta_W = \pm 1$.  By identifying
	$\sin \theta_{\mu \nu}/(e a^2)$ as the field strength we then
	obtain an Ehrenfest theorem of the form:
\ben
\la
\frac{1}{a}\Delta_{\mu} F_{\mu \nu} - J_{\nu}^{\lab{(static)} } 
\ra_{\lab{source} }  =  0.  
\een
	The choice of $\theta_{\mu \nu}/(e a^2)$ as the field strength
	would not lead to an Ehrenfest theorem for finite lattice
	spacing.

\subsection*{\bf Generalization to SU(2)}

	The SU(2) Wilson plaquette action, $S$, gives
\ben
{\cal Z}_{W} = \int [d U_{\mu}] e^{i \theta_W}
\exp{\left(- \beta  S \right)},
\een
	where $W$ now indicates an Abelian Wilson loop. Ignoring gauge
	invariance for the moment, we exploit the invariance of the
	measure under a right (or left) multiplication of a link
	variable by a constant SU(2) matrix,
$
U_{\mu} \rightarrow 
U_{\mu} \left(1 - \frac{i}{2} \epsilon \sigma_3 \right).
$
	The derivative of $S$ with respect to $\epsilon$, which we
	denote as $S_{\mu}$, inserts a $\sigma_3$ in the six
	plaquettes contiguous to the shifted link.  Similarly the
	Abelian Wilson loop has a $\sigma_3$ insertion if it contains
	the shifted link.
	This gives the relation
\be
\la \frac{2}{g a^3} S_{\nu}(U) - J_{\nu}^{\lab{(static)} } 
\ra_{\lab{Abelian source} } = 0,
\label{eqn_ap_nogf}
\ee

	This can be cast into the form Eqn.~(\ref{eqn_ehren}) using
	the parametrization of the link matrix:
\be
U_\mu = 
 \left (
\begin{array}{cc}
\cos (\phi_\mu ) e^{{\textstyle i\theta_\mu}} &
\sin (\phi_\mu ) e^{{\textstyle i\chi_\mu }} \\
-\sin (\phi_\mu ) e^{{\textstyle -i\chi_\mu }} &
\cos (\phi_\mu ) e^{{\textstyle -i\theta_\mu }} \\
\end{array}
\right ), 
\label{eqn_decomp}
\ee
	where $\phi_\mu \in [0,\frac{\pi}{2})$ and $\theta_\mu$,
	$\chi_\mu \in [-\pi,\pi)$.
	Separate this into the diagonal, $ {\cal D}_{\mu}$, and
	off-diagonal, $ {\cal O}_{\mu}$, parts;
$
 U_\mu  = {\cal D}_{\mu} + {\cal O}_{\mu}.
$
	Applying this to 
	Eqn.~(\ref{eqn_ap_nogf}) gives
\be
\la  \frac{1}{a}\Delta_{\mu} F_{\mu \nu} -
 J_{\nu}^{\lab{(dynamical)} } - J_{\nu}^{\lab{(static)} } \ra = 0 ,
\label{eqn_ehren_nogf}
\ee
	where $\frac{1}{a}\Delta_{\mu} F_{\mu \nu}$ contains only
	${\cal D}_{\mu}$ contributions to the links and
	$J_{\nu}^{\lab{(dynamical)} }$ the rest.

\subsection*{\bf Gauge fixing}

	Gauge fixing complicates the issue, although the essence of
	the argument goes through as before. Prior to Abelian
	projection we gauge fix to satisfy $F_i (U; n) = 0$ for
	$i=1,2$. When shifting a link 
$ 
U_{\mu} \rightarrow 
U_{\mu} \left(1 - \frac{i}{2} \epsilon \sigma_3 \right) 
$
	we must in general perform a simultaneous gauge
	transformation,
$
g(n) = 1 - \frac{i}{2} \mb{\eta } \cdot \mb{\sigma } $,
	where $\eta^i(n) \propto \epsilon$, to avoid leaving the gauge
	condition. $S_\mu$ is gauge invariant, but we obtain extra
	terms from the Wilson loop source and the Faddeev--Popov
	determinant when we differentiate with respect to $\epsilon$.

	The partition function now reads:
\ben
{\cal Z}_{W} = \int [d U_{\mu}]\; e^{i \theta_W}\; \Delta_{FP} \;
 \smprod_{n,i}  \delta(F_i (U; n)) \;
e^{-\beta  S }.
\een
	 We are primarily interested in the MA gauge,
\ben
F_i (U;n) 
&=& \frac{1}{2}\sum_{ \mu} 
\left\{
\hbox{Tr}
\left(
\sigma_i U^{\dagger}_{\mu}(n) \sigma_3 U_{\mu}(n)
\right)
\right.
\\ &+& 
\left. 
\hbox{Tr}
\left(
\sigma_i U_{\mu}(n - \hat{\mu}) \sigma_3 
U_{\mu}^{\dagger}(n - \hat{\mu})
\right) 
\right\},
\een
	and integrating out the ghost fields gives
\ben
\Delta_{FP} = \det 
\left( \frac{\partial F_i(U;n)}{\partial \eta_j(m)}\right)_{F_i (U;n)=0}.
\een
	The Ehrenfest theorem is now given by
\ben
 \int [d U_{\mu}]\; e^{i \theta_W}\; \Delta_{FP} \;
 \smprod_{n,i}  \delta(F_i (U; n)) \times \\ 
 \left(- \beta S_{\mu}  + 
 \frac{(\Delta_{FP})_\mu}{\Delta_{FP}} + i (\theta_{W})_\mu \right)
e^{-\beta  S } = 0.
\een
	which is recast as Eqn.~(\ref{eqn_ehren}).


%
%
%
%
%
\subsection*{\bf Status and Conclusions}

	When shifting the link does not conflict with the gauge
	condition, e.g. when no gauge condition is imposed, no extra
	gauge transformation is required and we find
	Eqn.~(\ref{eqn_ehren_nogf}) is satisfied exactly.

	Ignoring this conflict in the MA gauge, in a normalization
	where $ \langle J_\nu^{\lab{(static)} } \rangle= 1$ on the
	source the sum of the first two terms in
	Eqn.~(\ref{eqn_ehren_nogf}) is $1.128$~$(5)$. We have used a
	Abelian plaquette source ($\beta = 2.3$ on a $12^4$
	lattice). Hence there is a $13\%$ violation of the
	identity. On the same lattice, in the numerically simpler
	gauge that diagonalizes all plaquettes in a particular plane,
	the violation is $-23\%$. In both cases there is a rapidly
	decreasing, but non-zero signal for the summed terms that
	extends away from the source where $\langle
	J_\nu^{\lab{(static)} } \rangle = 0$.

	The corrective gauge transformation at every site that
	accompanies the shift of a single link introduces a
	non-locality. The Wilson loop derivative, $(\theta_W)_\mu$, is
	increased, and picks up a contribution even when the shifted
	link is not one of those making up the loop. The magnitude of
	the gauge transformation falls off exponentially with
	distance, however, and is a small effect. In other words, the
	derivative of the source is no longer a delta function of
	position, but is slightly smeared by the gauge fixing.

	Inclusion of this reduces the violation of the MA gauge
	Ehrenfest identity to $3\%$ on a $6^4$ lattice at $\beta =
	2.3$ (since this is a lattice identity, finite volume and
	scaling considerations are irrelevant). On such a lattice, the
	plaquette gauge relation is improved to $-15\%$.

	The calculation of the Faddeev--Popov term is incomplete, but
	preliminary results indicate that $\langle
	J_\nu^{\lab{(Faddeev--Popov)} } \rangle$ is relatively small
	and of the correct sign and magnitude to cancel the remaining
	violation and satisfy Eqn.~(\ref{eqn_ehren}) both in the MA
	and plaquette gauges.

	The lattice definition of Abelian field strength that follows
	from this approach is
\ben
F_{\mu \nu}(n) = \frac{1}{g a^2} 
\times \hspace{4.2cm} \\
\hbox{Tr }  \left[\sigma_3 {\cal D}_{\mu}(n)
{\cal D}_{\nu}(n + \hat{\mu})
{\cal D}^{\dagger}_{\mu}(n + \hat{\nu})
{\cal D}^{\dagger}_{\nu}(n)\right].
\een
	This differs from the lattice version
\cite{bdh}
	of the 't Hooft Abelian field strength
\cite{thooft2}.
	Both agree in the continuum limit, but the latter would
	satisfy the Ehrenfest theorem only to leading order in $a$.

	In terms of the link parametrization, Eqn.~(\ref{eqn_decomp}),
	the field strength is:
\ben
F_{\mu \nu}(n) = \frac{1}{g a^2} 
\cos \theta_{\mu \nu}(n) \times \left\{ \cos \phi_{\mu}(n) 
\right.
\hspace{4em}\\
\left.
\cos \phi_{\nu}(n+\hat{\mu}) 
\cos \phi_{\mu}(n+\hat{\nu})  \cos \phi_{\nu}(n) 
\right \}.
\een
	The factors $ \cos \phi_{\mu}(n) \rightarrow 1$ in the
	continuum limit.

\end{document}